# A Muti-channel Distributed DAQ for n-TPC*

Cheng Xiaolei（程晓磊）[1], Liu jianfang（刘建芳）[1], Yu Qian（余谦）[1] Niu libo（牛莉博）[2], Li Yulan（李玉兰）[2]

[1] School of Nuclear Science and Engineering, North China Electric Power Univ., Beijing 102206, China

[2] Department of Engineering Physics, Tsinghua University, Beijing 100084, China

**Abstract** A new fast neutron spectrometer named n-TPC has been designed by LPRI (Key Laboratory of Particle & Radiation Imaging, Ministry of Education) at Tsinghua University. The neutron energy spectrum can be calculated from the recoil angle and energy of the recoil proton detected by a 704-pad GEM-TPC. In beam tests at IHIP (Institute of Heavy Ion Physics, Peking University) in 2014, n-TPC performed better than 6%@6MeV energy resolution and 5‰ detection efficiency. To find the best working parameters (the component and proportion of the gas, the high voltage between each GEM layer, etc.) of the n-TPC and support its application in various conditions, a multichannel distributed DAQ has been design to read out the signals from the 704 channels. With over 25 Ms/s sampling rate and 12 bit resolution for each channel, it can record the time and amplitude information as well as traditional DAQs in the TPC application domain. The main design objective of this distributed DAQ, however, is more flexible parameter modulation and operation. It can support the n-TPC without the limitation of the chassis and categorize signals arriving from the 704 channels at the same time by different events without event triggers.

**Keywords** data acquisition systems, distributed control, n-TPC, signals categorized by event, zero compression

**PACS** 29.85.Ca

## 1 Introduction

A 704-channel Time Projection Chamber (TPC) has been designed as an advanced fast neutron spectrometer [1] as shown in Fig. 1. Argon-hydrocarbon mixture is selected as the working gas for: 1) the convertor for converting neutrons to protons; 2) the TPC drift gas and 3) the GEM multiplication gas. As the system is symmetric about the incident axis, the readout pads are designed as arc segments around the axis as shown in Fig. 2. The recoil angle and energy deposition $E_p$ of the recoil proton can be obtrained from the fast neutron. The energy of the neutron $E_n$ can then be calculated by the formula $E_n=E_p*\cos^2\theta$ [1, 2].

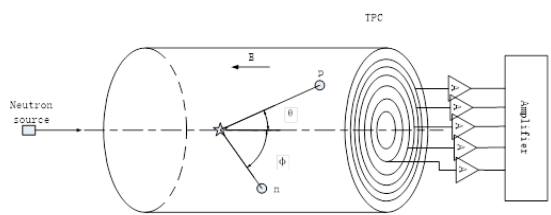

Fig. 1. Diagram of fast neutron spectrometer.

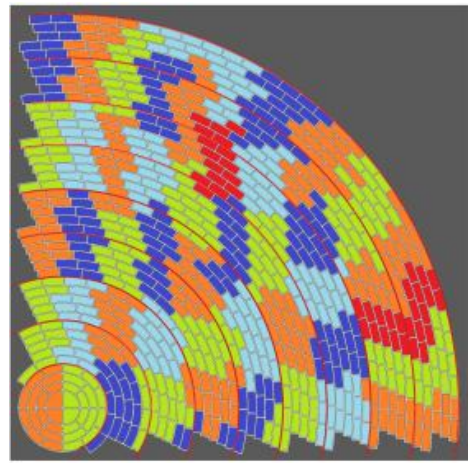

Fig. 2. Pad distribution of n-TPC. Different colors indicate different grouping of the pads to readout electronics.

Compared to the traditional spectrometer based on the proton-recoil method, this method has the following merits: 1) the working gas is used both as the convertor and the detection medium of the recoiled proton, avoiding energy loss in the convertor; 2) large coverage of recoil angle measurement, which can improve the detection efficiency; and 3) very good discrimination ability for protons from gamma

*Project supported by the National Natural Science Foundation of China (Grant No. 11275109).

or other recoil nuclei based on their very different *dE/dx* values [2].

A readout electronics system and customized DAQ is needed to obtain the drift time and the total ionization charge from the signal of each pad. Then the $E_p$ can be calculated by the sum of charge and θ can be calculated by the track projections on the pads plane and drift direction. The signals from n-TPC are processed by amplifier and pulse shaped circuits, and their bandwidths are restricted to less than 8 MHz. A 25 Ms/s sampling rate is fast enough to record the information of the signals output by the shaping circuits. Considering the calculations from LPRI, the dynamic range of each channel should be better than 2000 to achieve the 5% energy resolution for fast neutrons. The sampling rate and resolution of each DAQ channel are then designed as 25 Ms/s and 12 bits (dynamic range of nearly 4000).

There are two challenges, however, in which this DAQ differs from traditional DAQs for TPCs: 1) the form factor of this DAQ should be modulated flexibly to extend the application of the n-TPC in various conditions; 2) it is hard to get the start time of a nuclear recoil effect, so the 704-channel signals need to be categorized by different events without event triggers.

## 2 The distributed DAQ

As shown in TABLE I, a centralized system with bus protocol and chassis architecture will usually be the first choice for the DAQ of TPC [3-9]. The n-TPC, however, needs a DAQ which can be operated with lower cost and higher flexibility than other fast neutron spectrometers to demonstrate its superiority in energy resolution and detection efficiency. In addition, a chassis DAQ would cause demanding requirements on the installation location and space, and make it hard to achieve plug and play for new channels when the whole detection system is operating. For the objective to extend the application of this n-TPC, a distributed DAQ with independent power supply and uplink channel for each board should be best choice.

TABLE I
SUMMARY OF DAQs FOR TPCs

| Experiment & Equipment | DAQ Principle | DAQ Architecture |
|---|---|---|
| T2K | 25Ms/s 12bits FADC | MIDAS[b] based on CAMAC |
| ALICE TPC | 10Ms/s 10bits FADC | ALTRO[c] bus with chassis |
| PANDA TPC | 25Ms/s 12bits FADC | ALTRO bus with chassis |
| ILC GEM-TPC | 10Ms/s 10bits FADC | xTCA[d] being developed |
| ILC MM-TPC[a] | 25Ms/s 12bits FADC | xTCA being developed |

[a]MM=MicroMegas.
[b]MIDAS= Maximum Integration Data Acquisition System
[c]ALTRO=**AL**ICE **TPC R**ead **O**ut
[d]xTCA=ATCA(Advanced Telecommunications Computing Architecture) + microTCA + RTM(Rear Transition Module)

To meet the requirements of this n-TPC, we designed a distributed DAQ constructed with (n+1) boards as shown in Fig. 3. For the first version 704-channel n-TPC, this distributed DAQ supplies n=22 DSP boards with 32 A/D channels each to get the signals from the read out electronics (44 16-channel ASICs for the 704 pads of the n-TPC)[6], with a control board to synchronize the external trigger (if it can be provided), time reset and clock. This control board can ensure readout of the drift time with resolution as well as the sampling interval time of A/D. The time inconsistency between different boards can be reduced to the maximum. After zero compression and other processes, the sampling data with time scale are sent to a 22 port GigE switch through 100Mbps Ethernet interface on each board. Finally, all the data are sent to a recording computer through one 1Gbps network cable. All the boards of this DAQ are supplied by independent 5V3A power.

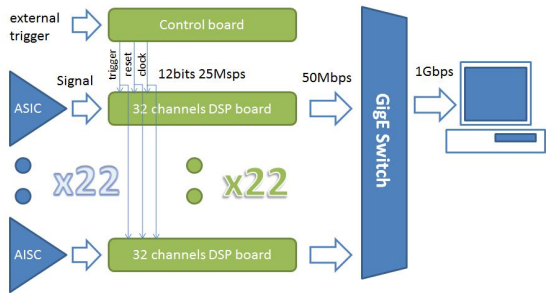

Fig. 3.  Design of the distributed DAQ.

This design can reduce the cost of a centralized power supply, which is over 10 times more expensive than the total cost of the independent power supplies. The number of boards can be changed flexibly with changes in the TPC readout pads, and the form factor of this distributed DAQ can be modulated flexibly for various installation conditions as shown in Fig. 4.

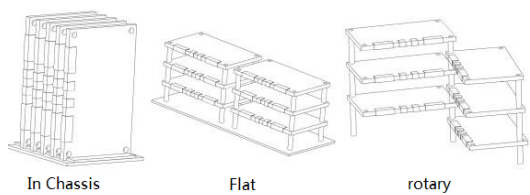

Fig. 4.  The DAQ can be installed in a 6U chassis if the installation space is sufficient. If there is not enough installation space, the DAQ can be installed flexibly and coordinated with the n-TPC to save space.

## 3 Functions of the DSP board

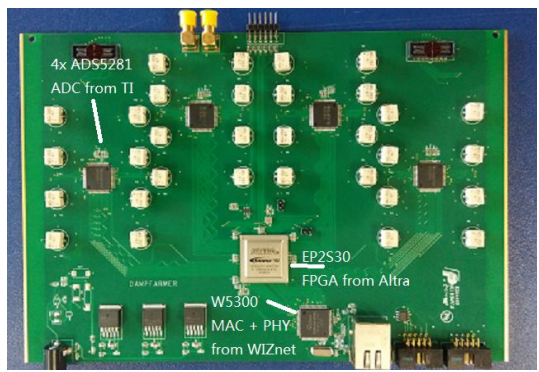

Fig. 5.  The DSP board. It can work independently without control and power supply from a carrier plate. All the intelligent functions are designed in one FPGA, without complex PCB logic circuits.

The DSP board is shown in Fig. 5. To meet the requirements for the n-TPC's energy resolution and time resolution [1], there are 4 8-channel ADC chips with 12 bit resolution and 25 Ms/s sampling rate on each board. All the sampling data are processed by a FPGA, in which we can select the trigger model of trigger, set the threshold of the self-trigger, finish zero compression and add the pulse arriving time scale to each data package. All those functions can be controlled online with a LabVIEW Program in the recording computer.

**3.1 Signals categorized without event trigger**

Even if the intensity of the fast neutron beam is stable, the time and location of the nuclear recoil effect and the recoil angle of the proton are all random. Without the accurate start and end time of a recoil proton flight, it is hard to distinguish which signals are generated by the same recoil event. Although we can record each random pulsed signal with a self-pre-trigger design in the FPGA, it is hard to get the relationship between all the signals without an external unified event trigger. Considering the theoretical recoil cross-section (less than 0.6%) of fast neutrons in an argon-hydrocarbon mixture, using a trigger generated by the neutron beam is not a good choice. Too much spurious triggering will waste system resources and reduce the detection efficiency.

To solve this problem, an intelligent function has been designed in the FPGAs. At power-on, the DAQ will start sampling the input signals and record data in a circular memory buffer (10 μs record length) for each channel. The old data will be overwritten sequentially by new data if the memory is full. If any channel is triggered by the first electronic cluster to arrive from one event, a 6.4 μs time window will be opened to wait for the last cluster to arrive in other channels. At the end of the 6.4 μs window, the DAQ will stop writing to the circular memory. The information coming before the trigger arrival is recorded in the 3.6 μs length old data in each channel, which will be uplinked to the computer with the data recorded in the 6.4 μs window.

All the channels triggered in 6.4 μs (which is the optimal time length to acquire all the signals from one event while effectively excluding accidental events) will be seen as the pads hit by the clusters generated by one recoil proton. All the data from the triggered channel can be uplinked immediately after the time window. The DAQ will then go back to the waiting state until the next recoil event. Through this function the acquired data will be categorized by event, and the recoil angle and the recoil proton energy will be calculated reliably.

In the n-TPC chamber, the drift speed of the electrons is about 4 - 7 cm/μs and the flight speed of the recoil protons is nearly $10^4$ cm/μs ($10^8$ m/s, the same order of magnitude as light). Compared to the electron drift time, the time between the generation and termination of a recoil proton is tiny enough to be ignored. Although the generation of a recoil proton is several picoseconds or femtoseconds earlier than its first ionization, its impact on the time resolution of the n-TPC is minimal.

**3.2 Zero compression**

The number of pads hit in one event can vary from 1 to 120, and the width of the pulsed signal from each pad can vary from 200 ns to 3 μs. This makes it difficult to improve the average event sampling rate of this DAQ. The maximum data size acquired from one event is 3000 ns×704× 12 bits/40 ns=633.6 kbits. Limited by the cost and interface technology, those data should be transmitted to computer over a single GigE Ethernet cable. The time of data uplink will last over 1 ms and the event sampling rate will be limited to less than 1 k/s.

In fact, most of the 1,013,760 data bits are '0'. Thus an intelligent function has been designed to ensure that only the 10 ns baseline data before the rising-edge and the 10 ns baseline data after the falling-edge of a pulsed signal will be uplinked together with the sampling data of the pulsed signal; other baseline data from the 1 - 120 triggered channels and the data from non-triggered channels will be ignored. Then the 'non-zero' data of the 1 - 120 triggered channels in the 6.4 ns time window will be transmitted to the computer, and the 'zero' data from all channels will be ignored. Calculated with the probability distribution of the recoil angles in the chamber as shown in Fig. 1, the average number of hit pads per event is nearly 72, and the average width of pulsed signals is nearly 1.24 us. The average data size acquired per event is then 26.784 kbits, and then the event sampling rate becomes over 20 k/s.

## 4 Application to n-TPC calibration

### 4.1 Operating parameter research

The optimal gas composition and bias voltage have to be tested to improve the energy resolution and spatial resolution of the n-TPC. A Mini-X-OEM product from AMPTEK is used to generate 8.04 keV X-rays to calibrate the operating parameters with different gas compositions and bias voltages.

The energy of the X-rays can be calculated from the sum of the charge collected by the hit pads in one event. The 2-D projection of the X-ray track on the plane of pads can be calculated by the number of pads hit and the distance between each pad. The projection in the perpendicular direction (relative to the plane of the pads), however, can be calculated by the difference in the charge drift time between the ionization at the two ends of the track (the drift speed is nearly constant in the n-TPC). The time consistency between each channel becomes the key influencing factor on the spatial resolution in the perpendicular direction. As shown in Fig. 6, the time difference between the two test channels becomes more serious as time goes on.

A synchronous clock is therefore generated by the control board and sent to each DSP board to synchronize each rising edge of the count clock for the time record counter in the FPGA, and then the time discretization between each channel is reduced to less than one sampling interval, as shown in Fig. 7.

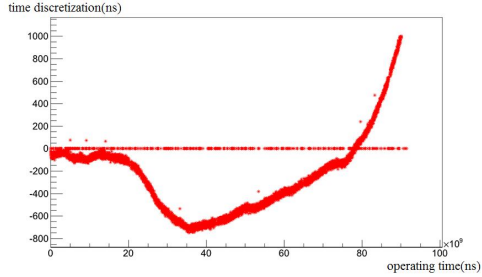

Fig.6. Time discretization in 100 s without synchronization.

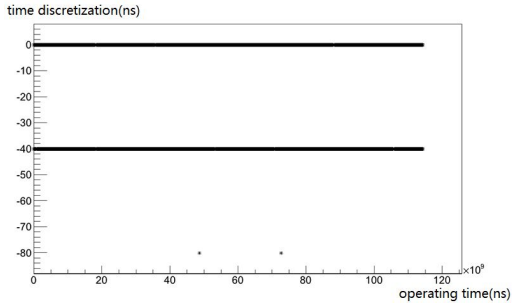

Fig.7. Time discretization in 100 s with synchronization.

Then, this distributed DAQ can record the pulsed signals from the n-TPC with sufficient resolution and channel consistency to test the detector's operating parameters as shown in Fig. 8. When changing the relative percentage of $Ar_2$ and $C_2H_6$, we found that the gain linearity became worse with increasing percentage of $C_2H_6$. If we reduce the percentage of $C_2H_6$, however, the probability of the nuclear recoil effect happening decreases rapidly. A gas ratio of 50:50 is therefore chosen for equilibrium.

As shown in Table II, a series of experiments proved that the energy resolution varies noticeably if the bias voltage on the single layer GEM changes. The optimal bias voltage is therefore fixed at 395 V.

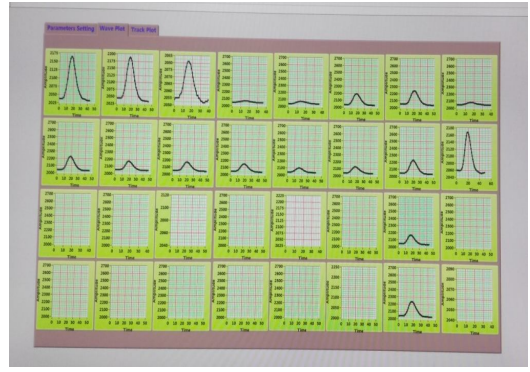

Fig. 8. The recording and controlling LabVIEW program. The pulsed signals are recorded channel by channel and shown in a group of 32.

TABLE II

8.04 KeV X-RAYS ENERGY RESOLUTION WITH DIFFERENT BIAS VOLTAGE

| Bias voltage | Operating state | FWHM |
|---|---|---|
| 375V | Signals inundated by noise | -- |
| 380V | normal | 45.2% |
| 385V | normal | 34.1% |
| 390V | normal | 30.7% |
| 395V | normal | 29.0% |
| 400V | Frequently sparking | -- |

**4.2 Test with proton beams**

The proton energy resolution and recoil angular resolution of the n-TPC has to be calibrated before the application of neutron beams. Fig. 9 shows a test setup with 6 MeV proton beams which has been completed at IHIP. The angle of the perpendicular direction can be adjusted from 0 to 30 degrees in this test.

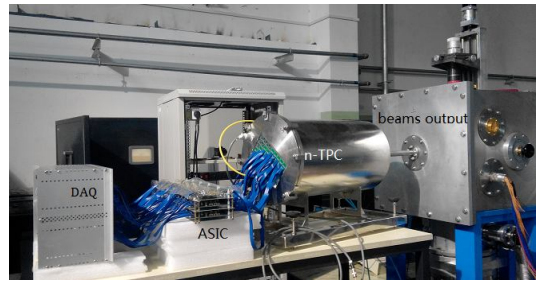

Fig. 9. Test setup with proton beams. The DAQ is assembled in a machine frame at the end of the test system.

As shown in Fig. 10, the 2-D and 3-D tracks of proton beams can be rebuilt by the DAQ.

With 11.2 sampling ENOB (Effective Number of Bits) and 120 ps max time discretization, this DAQ can support the efficiency and reliability of this n-TPC resolution test. As shown in TABLE III, the typical values of the proton energy resolution and recoil angle resolution of the n-TPC are calibrated. Limited by the space of the chamber, most protons flied out of the TPC when the incident angle was too large. These parameters verify the feasibility of advanced neutron spectrum measurement with the n-TPC.

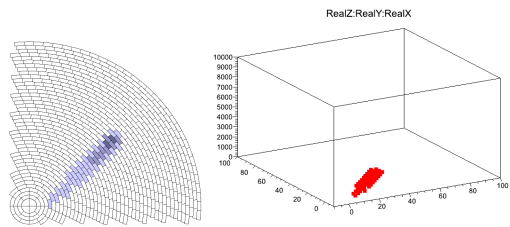

Fig. 10. Test result on 6MeV proton beams. The figure on the left is the 2-D projection of the proton track on the pads plane. The figure on the right is the 3-D rebuilding proton track.

The functions of categorizing data by event and zero compression are also validated in this test. The error rate of the data categorization is less than 0.3‰, and this DAQ can capture over 10000 proton tracks per second.

TABLE III

TEST RESULT OF PROTON ENERGY RESOLUTION AND RECOIL ANGLE RESOLUTION

| Proton energy (MeV) | Incident angle (degree) | Most of proton tracks | FWHM for recoil angle | FWHM for energy |
|---|---|---|---|---|
| 6 | 10 | In TPC | 3.95% | 5.3% |
| 6 | 20 | In TPC | 4.29% | 7.4% |
| 6 | 30 | Fly out | -- | -- |
| 6.5 | 10 | In TPC | 2.86% | 6.6% |
| 6.5 | 20 | In TPC | 3.77% | 7.7% |
| 7 | 10 | In TPC | 2.50% | 6.9% |
| 7 | 20 | In TPC | 2.54% | 7.9% |
| 7.5 | 10 | In TPC | 2.37% | 7.3% |
| 7.5 | 20 | Fly out | -- | -- |

## 5 Conclusions

The distributed DAQ can support the n-TPC for applications in various experiment conditions with its optimal form factor. Its ENOB and time consistency support the excellent energy resolution and angular resolution of the n-TPC. With intelligent functions designed in the core FPGA on each DSP board, the acquired data can be categorized by event without event triggers, and the sampling rate of events can reach over 10 k/s.

____________________________________